\def\rhob{\rho_{\rm b}}
\def\Pb{P_{\rm b}}

\documentclass{article}
\usepackage{graphics}
\input{epsf}
\begin{document}
\title{Avoided crossings in radial pulsations of neutron and strange stars}
\author{D. Gondek~~~~~J.L. Zdunik\\
N.Copernicus Astronomical Center, Polish Academy of Sciences,\\ Bartycka 18, PL-00-716 Warszawa, Poland\\
e-mail: dorota, jlz@camk.edu.pl}
\maketitle

\begin{abstract}
Radial pulsations of neutron stars and strange quark stars with
nuclear crust are studied.
The avoided crossing phenomenon occurring for the {\it radial} modes
is found and discussed. Neutron star models are constructed using
a realistic equation of state of dense matter and strange star
models using a phenomenological bag model of quark matter. The
eigenfrequencies of the three lowest modes of linear,
adiabatic pulsations are calculated, using the relativistic equations for
the radial oscillations.

\end{abstract}
%

\section{Introduction }
Radial oscillations of dense stars (neutron and strange stars)
were studied by many authors (see eg. Glass \& Lindblom
(\cite{GL}), V\"ath \& Chanmugam (\cite{VC}) and references
therein).

In the present paper we study the radial pulsations of neutron
stars and strange quark stars with nuclear crust in the framework of
general relativity. Our aim is
to present an interesting phenomenon
-- avoided crossing of this radial modes.
Avoided crossing manifest itself as a rapid change of the
frequency of the certain oscillatory mode when stellar
configuration gradually changes. At this point the frequencies of
two consecutive modes are very close and star "switches"  the
oscillatory properties from one to another. This effect was
previously noticed in the analysis of the {\it nonradial}
oscillations in "ordinary"  stars (Aizenman, Smeyers and Weigert
(\cite{ASW77}), Christensen-Dalsgaard (\cite{CD})) and has been
recently discussed by Buchler, Yecko and Koll\'ath (\cite{BYK}) in
the case of radial pulsations of the classical variable stars. For
neutron stars the avoided crossing phenomenon was considered by
Lee and Strohmayer (\cite{LeeS}) for {\it nonradial} oscillations
of rotating neutron stars.

The avoided crossings of radial modes in neutron stars was
mentioned in our recent paper (Gondek, Haensel and Zdunik
(\cite{GHZ}, hereafter Paper I) also in the case of hot
protoneutron star.

In this paper we study the avoided crossing phenomenon for two models
of dense matter: the low temperature limit of the
Lattimer and Swesty (\cite{LS91}) equation of state of the neutron matter and
the strange quark matter described by the simple bag model.

The plan of the paper is as follows. In Section 2 we describe
equations of state for which we found the avoided crossing
phenomenon. We calculate here also the adiabatic index of the
matter which is crucial for the existence of the avoided
crossings. In Section 3 we present the formulation of the problem
of linear, adiabatic, radial pulsations of stars. We discuss here
also the method of determination separately the oscillatory
properties of the inner part of a star and the outer regions
(envelope). Numerical results for the eigenfrequencies of the
lowest modes of radial pulsations of stars are presented in
Section 4. Finally, Section 5 contains a discussion of our results
and conclusions.
%


\section{Equation of state and adiabatic indices}

The starting point for the construction of our equation of state (EOS)
for the neutron
star models was the model of hot dense matter of Lattimer and
Swesty (\cite{LS91}), hereafter referred to as LS. Actually, we
used one specific LS model, corresponding to the incompressibility
modulus at the saturation density of symmetric nuclear matter
$K=220$~MeV. We use LS model of matter in the low temperature
limit, which can be treated as a zero-temperature $T=0$ model.

The detailed discussion of the EOS and equilibrium conditions is presented in
Paper I.

As a second example we consider the strange stars with nuclear
crust. The interior of this star is build of strange quark matter,
containing nearly equal number of $u$, $d$ and $s$ quarks. This
matter is described in the framework of the bag model with the
value of bag constant $B=60\ {\rm MeV\,fm^{-3}}$ (Witten
(\cite{Wit}), Farhi \& Jaffe (\cite{FJ}), Haensel, Zdunik \&
Schaeffer (\cite{HZS})). The crust of a quark star, at densities
below neutron drip point ($\rho_{\rm ND}=4.3\cdot 10^{11}\ {\rm
g\,cm^{-3}}$), is represented by  the BPS equation of state (Baym,
Pethick \& Sutherland \cite{BPS}).

\begin{figure}     
 \resizebox{\hsize}{!}{\includegraphics{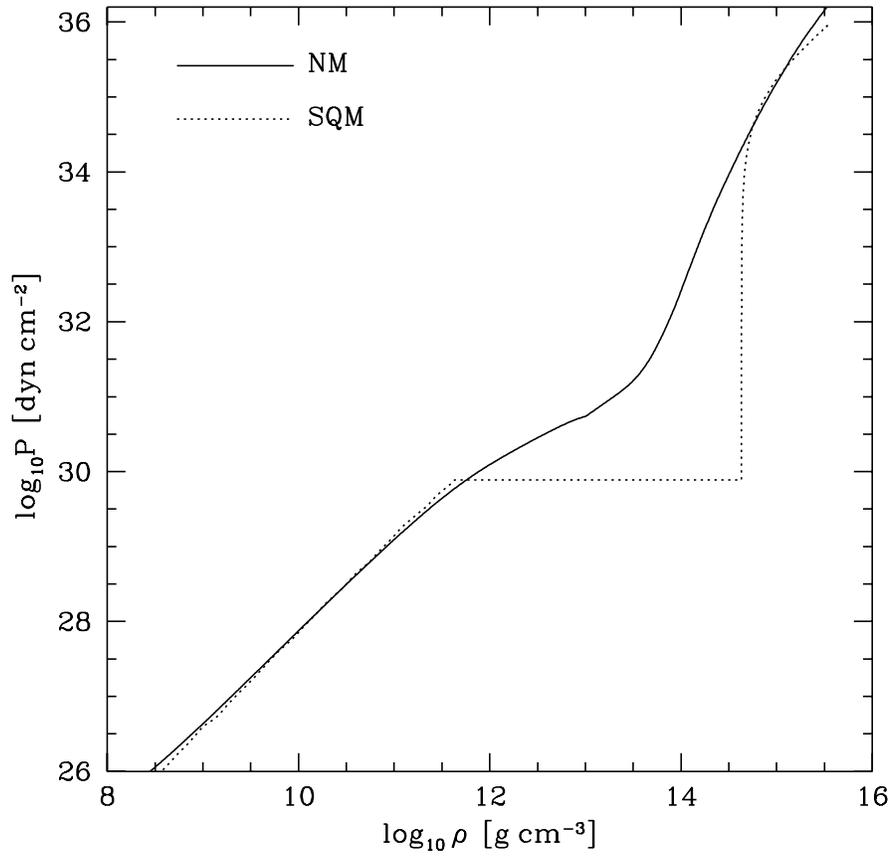}}
 \caption{Pressure versus baryon density for our models of dense matter.
The solid curve corresponds to cold catalyzed nuclear matter. The
dotted curve presents the EOS for strange quark star with nuclear
crust.}
\label{eos}
\end{figure}

In Fig.~\ref{eos} we show our equations of state for nuclear and quark matter.

\subsection{Adiabatic indices}
The adiabatic index within the star, $\Gamma$, plays  a crucial
role for oscillatory properties of the star.

The oscillation frequency depends sensitively on the value of the
adiabatic index and on the shape of the $\Gamma(\rho)$ function.
The increase of $\Gamma$  means the growing stiffness of the
matter leading to the effect of the "wall", setting the bounds for
the regions of oscillatory motion.

To calculate properly the oscillations of the star one has to know the
relevant index, governing linear perturbation of the pressure due to the
density perturbation. This index will be denoted by
$\Gamma_{\rm osc}$.

In the case of a sufficiently cold matter, the reactions between
matter
constituents are so slow, that the matter composition remains
fixed (frozen) during perturbations in the dynamical timescale, because
$\tau_{\rm react}\gg \tau_{\rm dyn}$.
In such a case, the
appropriate adiabatic index is $\Gamma_{\rm frozen}$ i.e.
\begin{equation}
\Gamma_{\rm osc}=\Gamma_{\rm frozen} \equiv
      {n\over P}\left({{\rm d}P\over {\rm d}n}\right)_{s,Y_e}~,
\label{Gamma_frozen}
\end{equation}
where $s,Y_e$ correspond to the equilibrium model. Fixed value of
$Y_e$ during oscillatory motion results in freezing of the weak
interaction processes ($\beta$-reactions).

The detailed discussion of the relevant adiabatic indices $\Gamma_{\rm osc}$
at different physical conditions (hot protoneutron star with or without
trapped neutrinos) was presented in Paper I.

\begin{figure}     
\leavevmode \resizebox{\hsize}{!}{\includegraphics{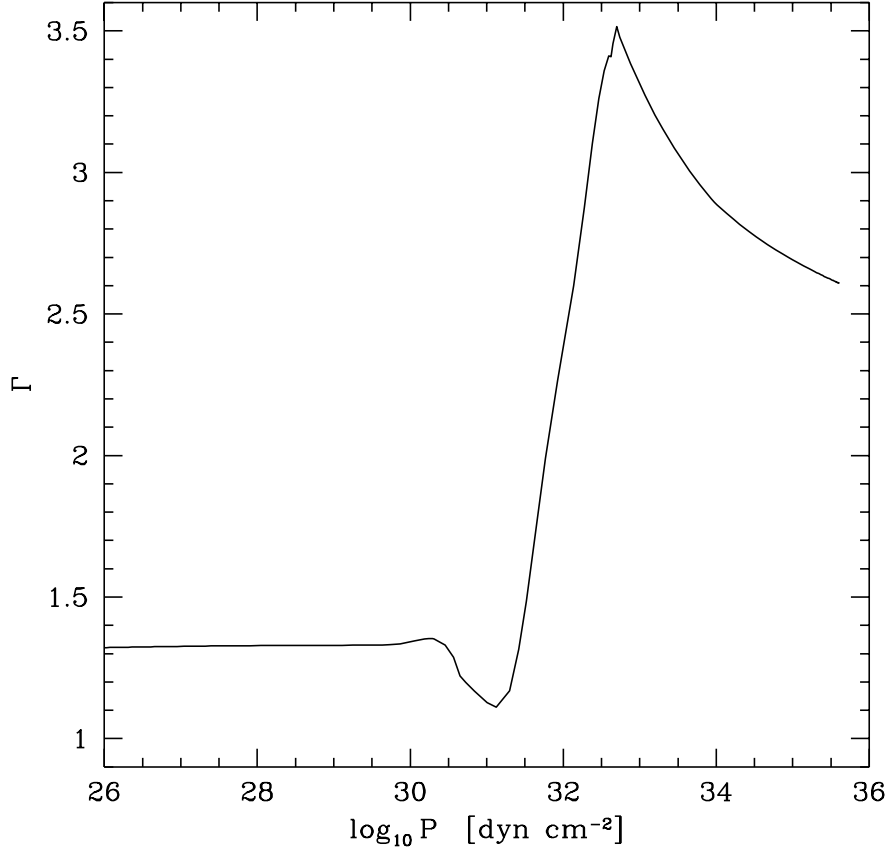}}
\caption{Parameter $\Gamma_{\rm osc}$ versus pressure, for the
nucleon matter.}
\label{gamn}
\end{figure}
\begin{figure}     
 \resizebox{\hsize}{!}{\includegraphics{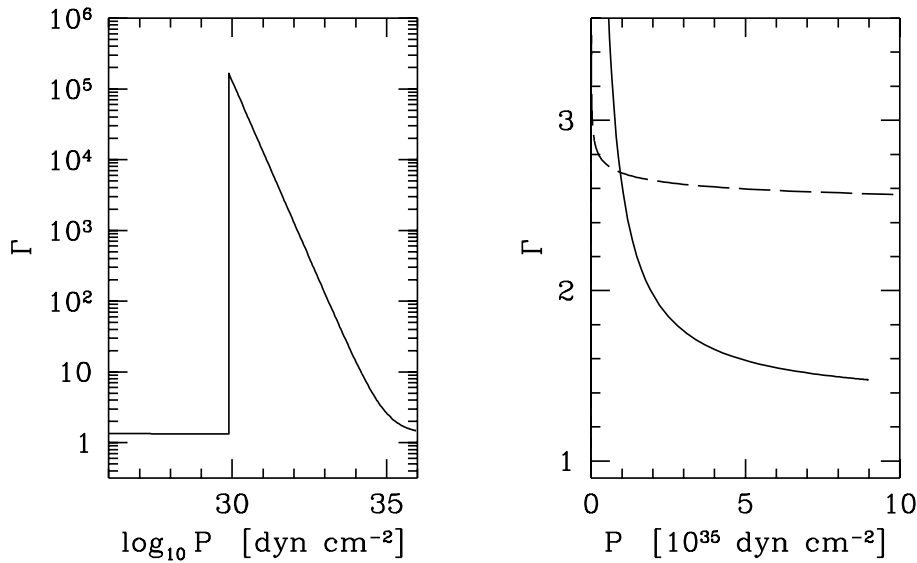}}
\caption{Parameter $\Gamma_{\rm osc}$ versus pressure, for the
strange quark matter. The left figure in logarithmic scale
visualizes the huge jump in $\Gamma$ at the crust--strange core
boundary. In the right figure we presented $\Gamma$ in linear
scale in the region of core oscillations. For comparison we show
$\Gamma$ for nuclear matter (dashed line) }
\label{gamq}
\end{figure}

The dependence of $\Gamma$'s on the pressure in the stellar
interior, for our models  is displayed in
Figs.~\ref{gamn},~\ref{gamq}. In this figures (and also some next
figures) we choose pressure as an independent variable because
pressure is continuous through the star and describes very well
some stellar parameters (e.g. mass above given radius). This is
not the case of density $\rho$, which exhibits large density jump
at the crust--core boundary of quark stars.
%

\begin{figure}     
 \resizebox{\hsize}{!}{\includegraphics{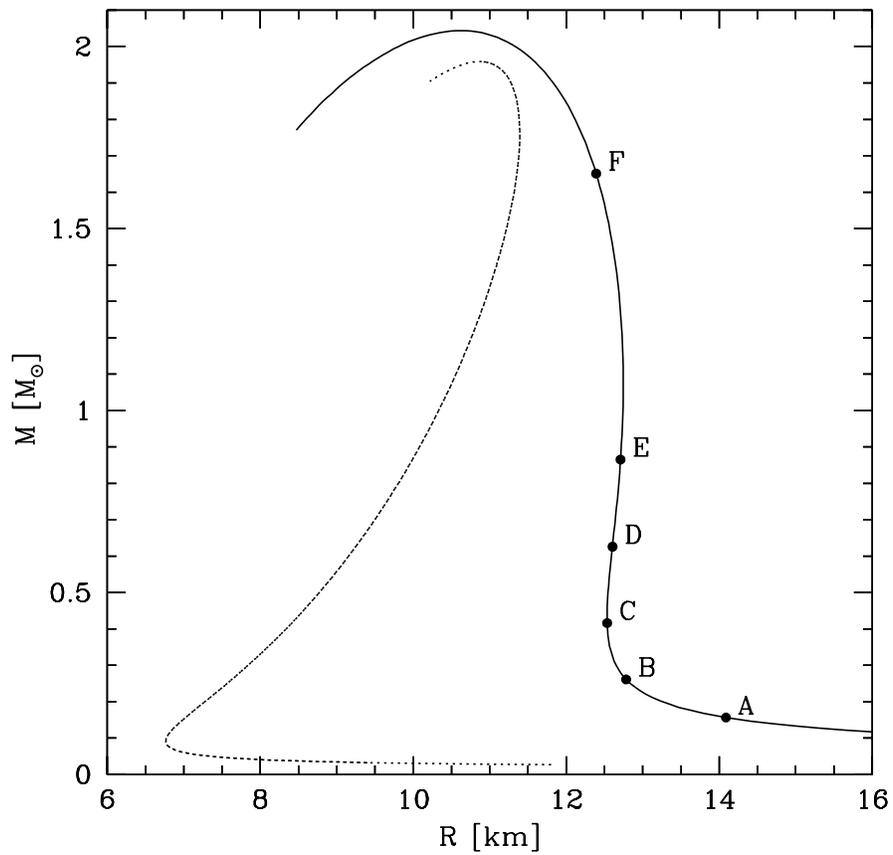}}
\caption{The gravitational mass versus stellar radius for static
models of neutron stars (solid curve) and quark stars (dotted
curve). The points on the M(R) curve for neutron stars correspond
to the stellar models for which oscillatory properties were
studied in details and presented in Fig.~\ref{enerosc}.}
\label{massr}
\end{figure}

\begin{figure}     
 \resizebox{\hsize}{!}{\includegraphics{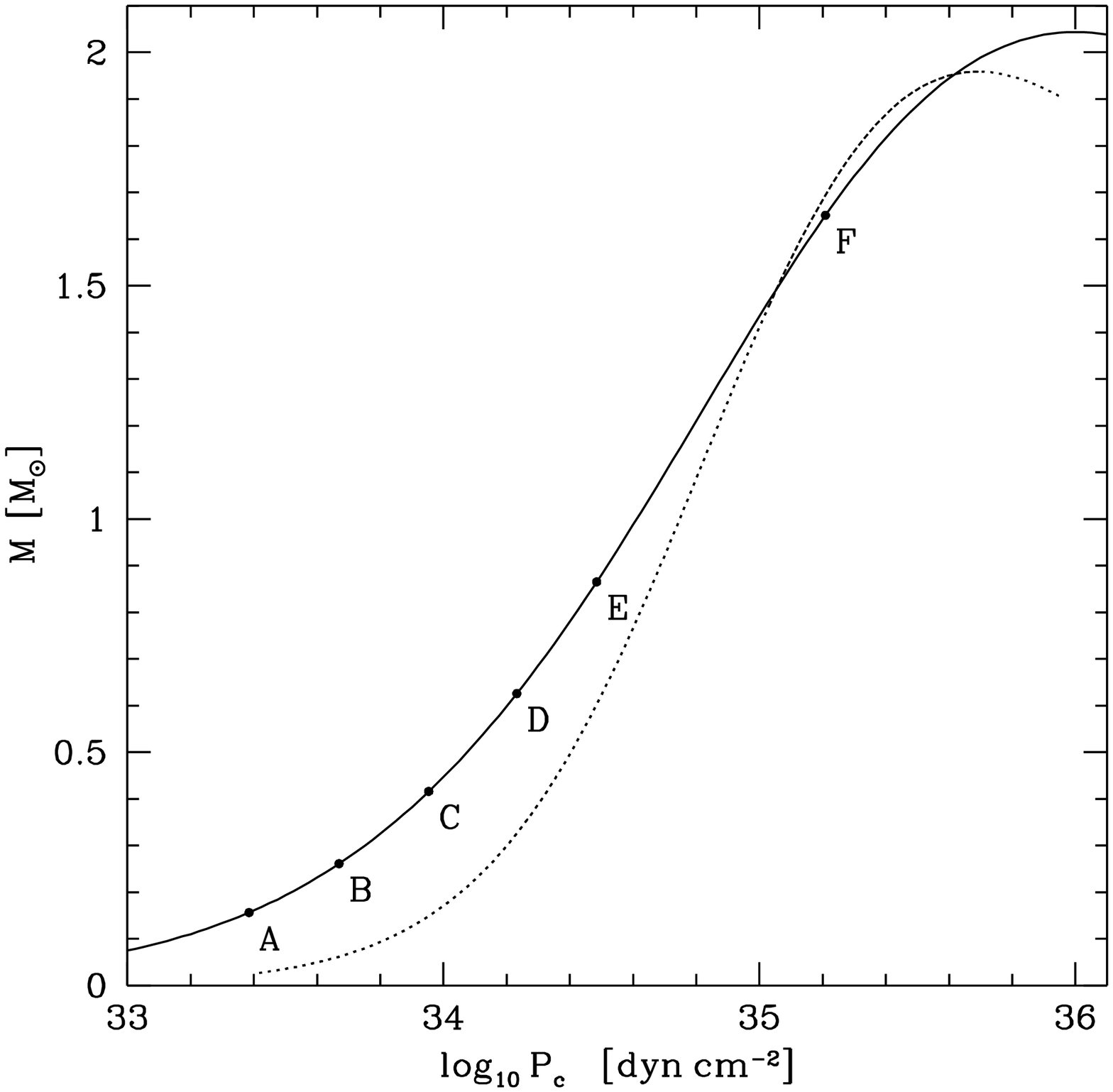}}
\caption{The gravitational mass versus central pressure for static
models of neutron stars (solid curve) and quark stars (dotted
curve). The points on the M(R) curve for neutron stars correspond
to the stellar models for which oscillatory properties were
studied in details and presented in Fig.~\ref{enerosc}.}
\label{masspc}
\end{figure}

\section{Linear adiabatic radial pulsations of stars}

The stellar configurations in hydrostatic equilibrium  are determined
by the integration of Tolman-Oppenheimer-Volkoff (TOV) equations (Tolman \cite{Tol},
Oppenheimer \&
Volkoff \cite{OV}).

The mass-radius relation for the neutron and quark star models
 is shown in Fig.~\ref{massr}.  In Fig.~\ref{masspc} we show mass vs.
 central pressure for our stellar configurations.
The points on the M(R) curve for neutron stars correspond to the
stellar models for which oscillatory properties were studied in
details and presented in Fig.~\ref{enerosc} and discussed in Sec.
4 .

To find eigenfrequencies $\omega$ we solve the equations governing
infinitesimal radial adiabatic
 stellar pulsations in general
relativity derived by Chandrasekhar (\cite{Chandra64}),
and rewritten by Chanmugan (\cite{Chanmugan77}) in a form, which
turns out to be particularly suitable for numerical
applications.
 Two important quantities, describing  pulsations, are:
 the relative radial displacement, $\xi = \Delta r/r$,
 where $\Delta r$ is the radial displacement of a matter element,
and $\Delta P$  -- the corresponding
  Lagrangian perturbation of the pressure. These two quantities
are determined from a system of two ordinary differential
equations which can be reduced
to the one second order (in $\xi$),
linear radial wave
equation, of the Sturm-Liouville type with $\omega^2$ as
the eigenvalue of the Sturm-Liouville problem.
As a result
for a given stellar configuration, we get a set of the
eigenvalues  $\omega_0^2 < \omega_1^2
<...<\omega_{\rm n}^2<...,$  with corresponding
eigenfunctions $\xi_0,\xi_1,...,
\xi_{\rm n},...,$ where the eigenfunction $\xi_{\rm n}$ has
$n$ nodes  within the star, $0\le r \le  R$ (see, e.g., Cox \cite{Cox80}).
The detailed form of the equations and solving method is presented in Paper I.

To solve linear radial wave
equation one needs two boundary conditions.
The condition of regularity at $r=0$ requires
\begin{equation}
\label{f}
\left(\Delta P\right)_{\rm center}=
-3\left(\xi \Gamma P\right)_{\rm center}~.
\label{bc1}
\end{equation}
 The surface of the star is determined by
 the condition that for  $r\rightarrow R$,
one has $P\rightarrow 0$. This implies
\begin{equation}
\label{g}
\left(\Delta P\right)_{\rm surface} =0
\label{bc2}
\end{equation}

To study the oscillations of the core end envelope separately we
divide the star into two parts: core from the center down to the
boundary density $\rhob$ and envelope with density smaller than
$\rhob$. In the case of neutron matter the density $\rhob$ is
defined by the steep change in $\Gamma$ (Fig. \ref{gamn}) and is
equal to $\simeq 10^{14}~{\rm g\,cm^{-3}}$. The corresponding
pressure is equal to  $\Pb\simeq 10^{32}~{\rm dyn\,cm^{-2}}$. For
quark stars the crust--core boundary is very well defined and
connected with huge density jump at the pressure $\Pb =
10^{30}~{\rm dyn\,cm^{-2}}$. At this boundary the density of the
crust is equal to $\rho_{\rm ND}=4.3\cdot 10^{11}\ {\rm g\,cm^{-3}}$
and the density of quark core is slightly above
$4\,B/c^2=4.28\cdot 10^{14}\ {\rm g\,cm^{-3}}$.

We have to apply the boundary conditions appropriate for core or
envelope pulsations. In the  investigation of the core
oscillations we stop the integrations of the linear radial wave
equation at the $\Pb$ with the boundary conditions corresponding
to the ''free'' envelope oscillation with the frequency defined by
the central core -- we treat the envelope as a mass $M_{\rm env}$
laying on the central core and simply moving in a way governed by
this central region. This assumptions leads to the boundary
condition in the form:

\begin{equation}
{\Delta P\over P}=-\left(4+\left({GM\over Rc^2}+{\omega^2R^3\over
 GM}\right)\!\left(1-{2GM\over Rc^2}\right)^{-1}\right)\xi
\label{bccore}
\end{equation}

The oscillation  of the envelope we can define by the boundary condition:
\begin{equation}
\xi(\rhob)=0
\label{bcenv}
\end{equation}
which means that central core do not move (in other words
$\xi(\rho\ge\rhob)=0$).

\section{Eigenfrequencies}

As a result of the numerical integration  of linear wave equation with
boundary conditions appropriate for three situations (Eqs. (\ref{bc1}, \ref{bc2},
\ref{bccore}, \ref{bcenv})) we obtain eigenfrequencies of oscillations for
each stellar configuration.
%
\begin{figure}     
 \resizebox{\hsize}{!}{\includegraphics{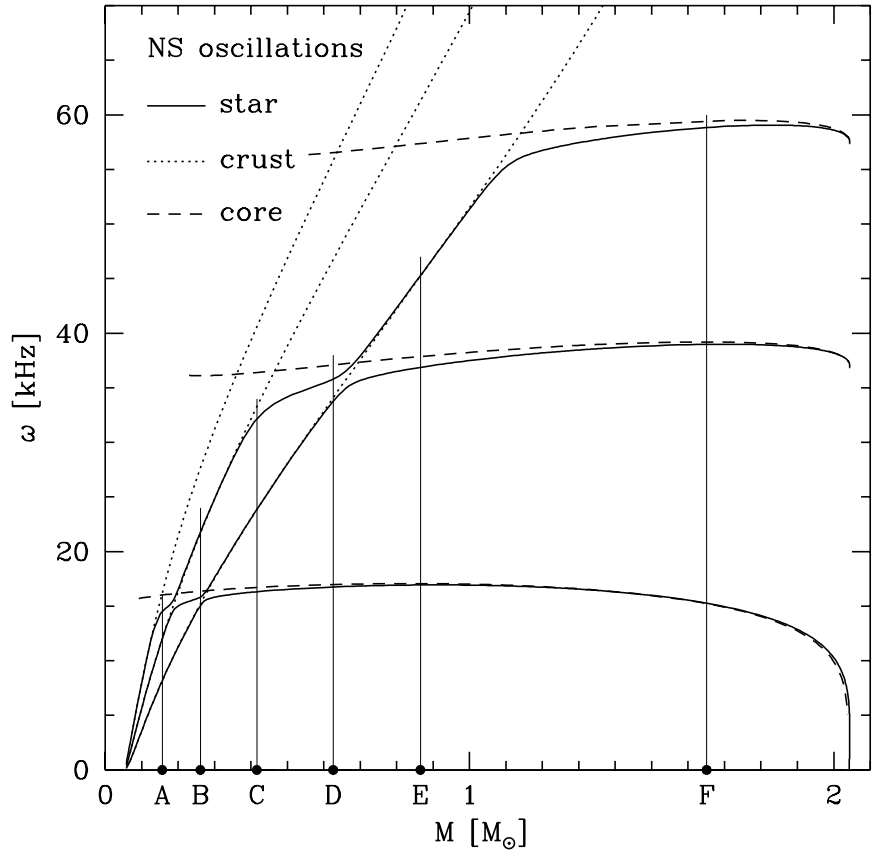}}
\caption{The eigenfrequencies of the three lowest modes, versus
stellar mass for neutron star's radial oscillations. The solid
curve corresponds to the oscillations of the star as a whole, the
dotted curve describes oscillations of the crust of a star and
dashed curve -- oscillations of the core. The points on the M(R)
curve for neutron stars correspond to the stellar models for which
oscillatory properties were studied in details and presented in
Fig.~\ref{enerosc}}
\label{omegan}
\end{figure}

\begin{figure}     
 \resizebox{\hsize}{!}{\includegraphics{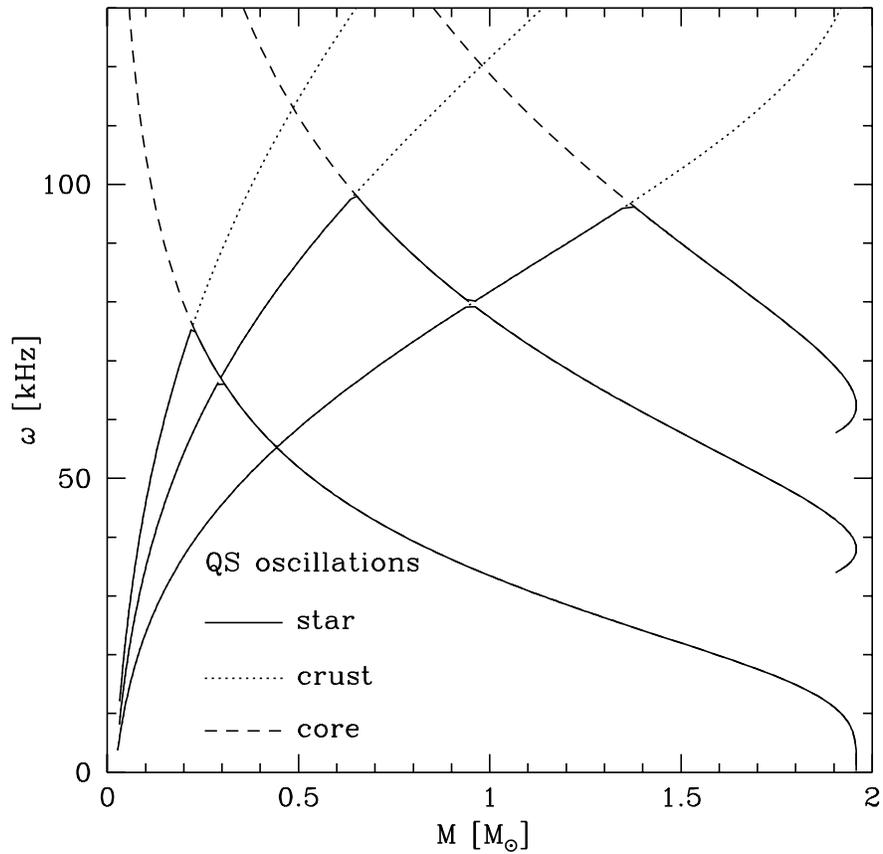}}
\caption{The eigenfrequencies of three lowest modes, versus
stellar mass for quark star's radial oscillations. The solid curve
corresponds to the oscillations of the star as a whole, the dotted
curve describes oscillations of the crust of a star and dashed
curve -- oscillations of the core. }
\label{omegaq}
\end{figure}

In Figs.~\ref{omegan}, \ref{omegaq} we plotted the eigenfrequencies of the three lowest
 $n=0,~1,~2$  radial modes versus stellar mass for neutron and quark
star oscillations respectively. The solid line corresponds to a
whole star pulsations with boundary conditions at the center and
stellar surface (Eqs. (\ref{bc1}, \ref{bc2})). The avoided crossing
phenomenon is clearly visible and manifest itself as abrupt,
nearly stepwise changes in frequencies of the consecutive
(neighbouring) oscillation modes. This effect occurs also in the
case of hot, isothermal models of protoneutron stars (Paper I).

This stepwise changes of $\omega_{\rm n}$  are due to the change
of the character of the standing-wave solution for the eigenproblem.
Namely, at the avoided crossing point the solution changes from the standing
wave localized mainly
in the outer layer of the star, to that localized predominantly  in the
central core.
This is very clearly visible from the comparison with solutions of oscillatory
equation for core and crust pulsations separately.
The solid line of the frequency of stellar pulsations nearly coincides with
either "core" or "crust" solutions. With increasing mass of the star
the subsequent modes $\omega_{\rm n}(M)$ choose between the crust or core solution to
 fulfill the condition  $\omega_0^2 < \omega_1^2
<...<\omega_{\rm n}^2<...,$. Because of the different dependence
$\omega_{\rm n}(M)$ for crust and core oscillatory modes this leads to
the avoided crossing effect.

The crust region in the stars under consideration is relatively
small. Thus the oscillations of the crust could be very well
described neglecting the mass of the crust in the gravitational
force. In the Newtonian limit and for constant $\Gamma=\Gamma_{\rm
crust}$ throughout the stellar crust the wave equation for crust
oscillations can be reduced to the form of Bessel equation of the
order of $\nu$ defined by the formula:
\begin{equation}
\Gamma=1+{1\over\nu}
\end{equation}
The resulting oscillatory frequency should then be
proportional to the gravity at the stellar surface
\begin{equation}
\omega_{\rm n}={1\over2}\,{j_{\nu, {\rm n}}\over
\nu\sqrt(1+{1\over\nu})}\, \sqrt{{\rhob\over \Pb}}\, g
\label{osccru}
\end{equation}
where $j_{\nu, {\rm n}}$ denotes the n-th zero of the Bessel
function $J_{\nu}(z)$ of the order of $\nu$ and $g={GM\over R^2}$
is gravitational acceleration at the stellar surface. This
approximation agrees very well with our exact numerical results
for neutron and quark stars. Especially for strange quark stars
where $\Gamma$ is nearly constant and equal to $4/3$ in the large
part of the crust (Fig. \ref{gamq}) the difference between true,
numerical results and approximation by the formula (\ref{osccru})
with $\nu=3$ is smaller than 5\%. In the case of neutron stars due
to the changes of $\Gamma$ through the outer layers the
identification of subsequent oscillatory modes with Bessel
functions  is less accurate, but the linear dependence of
$\omega_{\rm n}$ on $g$ in formula (\ref{osccru}) holds very well.

\begin{figure}     
 \resizebox{\hsize}{!}{\includegraphics{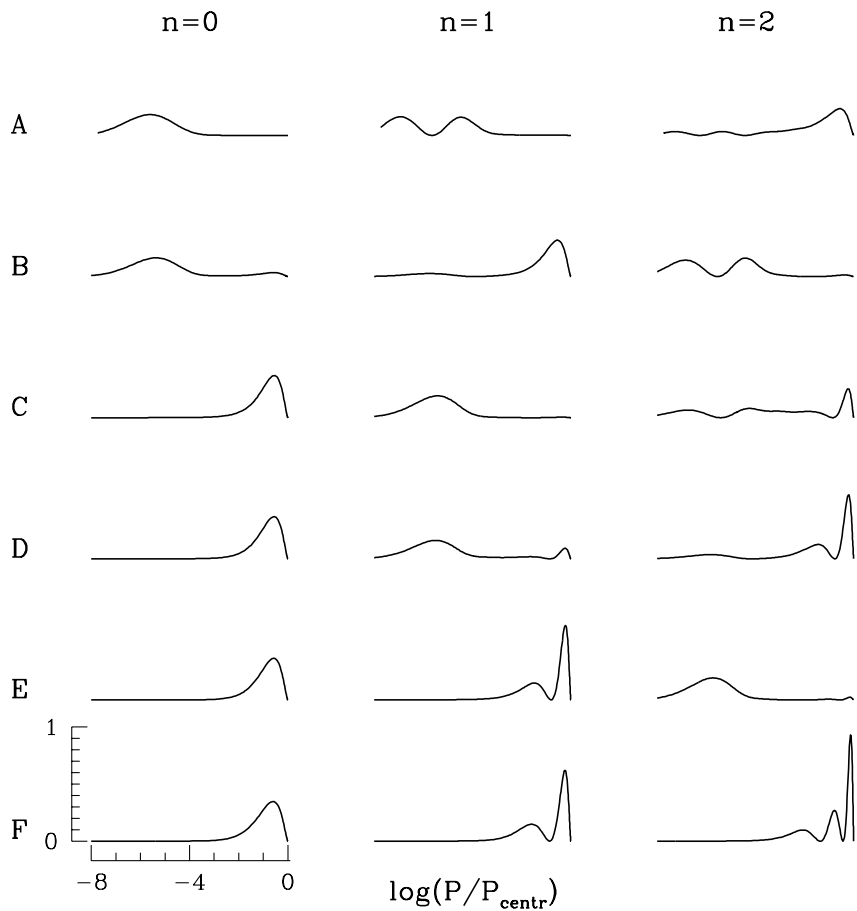}}
\caption{The oscillatory energy ($e_{\rm osc}$) vs relative
pressure ($P(r)/P_{\rm centr}$) for three lowest radial modes,
$n=0,~1,~2$ for neutron star. The subsequent rows (marked by
letters A--F) correspond to stellar models with growing central
density (from A to F) marked by dots in Figs. \ref{massr},
\ref{masspc}, \ref{omegan}. The energy is normalized to unity i.e.
$\int e_{\rm osc}\, {\rm d}(P/P_{\rm centr}) =1$. The function
$P(r)/P_{\rm centr}$ approximates very well the relative mass from
the stellar surface $(M-m(r))/M$. }
\label{enerosc}
\end{figure}

To confirm our conclusions we have studied in details properties
of the three lowest radial modes for some stellar configurations.
The parameters of this configurations and eigenfrequencies
$\omega_{\rm n}$ are presented in Figs. \ref{massr}--\ref{masspc}
by dots A--F. In Fig.~\ref{enerosc} we plotted the energy density
of oscillations calculated  per pressure gradient in the star i.e.
$e_{\rm osc}(P)\,{\rm d}P$ is equal to the energy due to the oscillatory
motion of the matter in the sphere $r(P)-r(P+{\rm d}P)$. This function
is proportional to $\xi^2_{\rm n}(r)$. As an independent variable
we choose in Fig.~\ref{enerosc} $P(r)/P_{\rm centr}$. For the
outer layers of the star this function is proportional to the
relative mass above the radius $r$ i.e.:
\begin{equation}
P(r)/P_{\rm centr} \sim (M-m(r))/M
\end{equation}
with proportionality constant of the order of unity.
The integral of the energy in Fig.~\ref{enerosc} is normalized to unity i.e.
$\int e_{\rm osc}\,{\rm d}(P/P_{\rm centr}) =1$.

The Fig.~\ref{enerosc} very well visualizes where the mode is
localized. We can also identify modes by the shape of $\xi$ or
energy of oscillation. The three lowest radial "core modes" are
characterized by the functions presented in row F in
Fig.~\ref{enerosc}. The stellar configuration F is the most
massive and crust plays very little role in oscillatory properties
(see also Fig.~\ref{omegan}). The two lowest "crust modes"
correspond to the columns $n=0$ and $n=1$ in row A of
Fig.~\ref{enerosc}. For the low massive model A the crust is
relatively large and determines frequencies for fundamental mode
and first overtone. Second overtone $n=2$ is strongly contaminated
by the core oscillations (Figs.~\ref{enerosc} and \ref{omegan}).
Fig.~\ref{enerosc} allows for very clear division of oscillatory
modes by the real pulsational properties and not the number of
nodes. This is presented in Table 1, where for example  A1 denotes
configuration A, number of nodes $n=1$.
\begin{table}
\caption{Identification of radial oscillatory modes of neutron star (see
Fig.~\ref{enerosc}) }
\begin{tabular}{llll}
\hline
\noalign{\smallskip}
propagation& fundamental&~~~first & second \\
region  & mode &~~~overtone &  overtone\\
\noalign{\smallskip}
\hline
\noalign{\smallskip}
crust&A0 B0 C1 E3&~~~A1 B2&\\
core&B1 C0 D0 E0 F0&~~~D2 E1 F1&~~~~F2\\
\noalign{\smallskip}
\hline
\noalign{\smallskip}
mixed&\multicolumn{3}{c}{A2 C2 D1}\\
\noalign{\smallskip}
\hline
\end{tabular}
\end{table}

The identification of the radial modes in Table 1 is consistent
with the Fig.~\ref{omegan}. The specific oscillatory mode for
stellar configurations A--F corresponds  to the crossection points
of the vertical lines at points A--F and the curves $\omega_{\rm n}(M)$
for given stellar configuration. The location of these points very
well define type of oscillatory motion. The nearly constant parts
of the $\omega_{\rm n}(M)$ functions correspond to the core oscillations
while regions of increasing $\omega_{\rm n}(M)$ describes crust
oscillations. The  region close to avoided crossing points is
characteristic to mixed modes: star pulsates in crust as well as
in the core.

\section{Discussion and conclusions}

The avoided crossing phenomenon in the {\it radial} pulsations of
dense stellar objects was presented. Both strange quark stars with
crust and neutron stars could be subject to this effect.

The main reason for the abrupt changes of the oscillatory
properties of the star as one change a little stellar
configuration is the very well defined division of the star into
two regions: inner core and outer crust. This two parts have
 different equation of state and oscillatory properties.
The latter have been studied in this paper by the separation of
the outer and inner regions by appropriate boundary conditions.
Resulting oscillatory frequencies of crust and core pulsations
depend on the stellar configurations (e.g. mass of the star) in
completely different manner. The functions $\omega_{\rm n}(M)_{\rm crust}$
are rather steeply increasing while $\omega_{\rm n}(M)_{\rm core}$ are
nearly constant (neutron stars) or decreasing (quark stars).

The oscillatory properties of the star could be characteristic
either to the crust or core pulsations. The frequency spectrum
depends on the mass of the star. The core pulsations determine the
oscillatory properties of the lower order radial modes for
relatively massive neutron or strange stars. The star could be
subject to the crust pulsations for sufficiently high frequency or
low stellar mass. For given mass the fundamental mode of crust
oscillations $\omega_0(M)_{\rm crust}$ define the frequency below
which the star pulsates in the core (Figs. \ref{omegan},
\ref{omegaq}).

The avoided crossing phenomenon is strongly related to the
changes of compressibility of the matter throughout the star,
which is described by the shape of the $\Gamma_{\rm osc}(\rho)$
function. The increase of the stiffness of the matter outward
leads to the maximum of $\Gamma_{\rm osc}(\rho)$ close to the
boundary between core and crust. In the case of strange quark
matter this is the consequence of the the fact that it is
self-bound at the very high density and $\Gamma\to\infty$ as
$P\to 0$. In neutron matter described by LS equation of state the
symmetry energy leads to the increase of $Y_e$ inward and in this
case more symmetric matter is softer. The same effect could be
seen for the field model of dense matter considered by Diaz Alonso and
Ib\'a\~nez
(\cite{DAI}). The function $\Gamma(\rho)$ for their EOSs
(Fig. 9 of their paper) resembles LS result at low temperature limit.

\vskip20pt

Acknowledgements

 We are very grateful to W. Dziembowski for helpful discussions.
This research was partially supported by the KBN grant No.
2P03D01413
and  by the KBN grant No. 2P03D01814 for D.
Gondek-Rosi\'nska.
 D. Gondek-Rosi\'nska was also supported by the program
R{\'e}seau Formation Recherche of the French Minist{\`e}re
de l'Enseignement Sup{\'e}rieure et de la Recherche.

\end{document}